\documentclass[reprint, superscriptaddress, amsmath, amssymb, aps, prl]{revtex4-2}
\usepackage{graphicx,dcolumn,bm,xcolor,etoolbox,blindtext,upgreek,float,nicefrac,mathrsfs,mhchem}
\usepackage[normalem]{ulem}
\usepackage[utf8]{inputenc}
\usepackage[OT1]{fontenc}

\definecolor{prlblue}{RGB}{46, 48, 146}
\definecolor{natcommblue}{RGB}{0, 104, 165}
\usepackage[colorlinks=true,citecolor=prlblue,urlcolor=prlblue,linkcolor=prlblue]{hyperref}
\usepackage{silence}
\begin{document}
\title{Spin-Lattice Dynamics and Interactions in Magnonic Spinels}

\author{Hari Paudyal}
\affiliation{Department of Physics and Astronomy, University of Iowa, Iowa City, Iowa 52242, USA}
\author{Yuri Suzuki}
\affiliation{Department of Applied Physics, Stanford University, Stanford, CA 94305, USA}
\author{Michael E. Flatt\'e}
\affiliation{Department of Physics and Astronomy, University of Iowa, Iowa City, Iowa 52242, USA}
\affiliation{Department of Applied Physics and Science Education, Eindhoven University of Technology, Eindhoven, The Netherlands}
\author{Durga Paudyal}
\affiliation{Department of Physics and Astronomy, University of Iowa, Iowa City, Iowa 52242, USA}

\date{\today}

\begin{abstract}
\noindent Minimizing magnetic damping while understanding spin-lattice interactions remains a key challenge for magnonics. We show site-specific Al/Li ordering in spinel ferrites drives a ferrimagnetic insulating state, quenching the Fermi-level density of states. \textit{Ab initio} calculations reveal collective acoustic modes alongside sub-lattice-selective optical modes on tetrahedral and octahedral networks, providing a microscopic understanding of substitution-driven spin dynamics. Crucially, the low-frequency magnon and acoustic phonon modes intersect, driving strong hybridization for low-loss technologies.
\end{abstract}

\maketitle

\noindent Harnessing spin waves in low-loss magnetic insulators with perpendicular magnetic anisotropy (PMA) offers a compelling pathway toward ultra-low-power, charge-free spintronic computing~\cite{wimmer2021mn, jungfleisch2016large, fert2017magnetic, bhatti2017spintronics}. Yet, identifying material systems that simultaneously combine robust ferromagnetic (FM) or ferrimagnetic (FIM) order, strong PMA, and minimal damping remains a critical bottleneck~\cite{demidov2015excitation, chumak2022roadmap}. Addressing this limitation requires a precise, microscopic understanding of fundamental exchange interactions, strain-driven lattice distortions, and magnon-phonon couplings. In this context, the magnetic damping parameter dictates the ultimate energy efficiency and propagation length in magnonic and spin-orbitronics. First-principles theory provides a powerful framework to unravel the intrinsic electronic and structural mechanisms driving dissipation across metallic, half-metallic, and insulating regimes. While the low damping in metallic ferromagnets is traditionally constrained by magnon scattering of conduction electrons, recent advances demonstrate that specific alloys of \text{Co} and \text{Fe} can exhibit an exceptionally low density of states (DOS) at the Fermi level. This unique electronic structure yields a damping parameter comparable to FIM insulators~\cite{schoen2016ultra}. Half-metallic FIM systems similarly suppress Gilbert damping by leveraging a unique spin-polarized DOS that fundamentally minimizes spin-flip and electron-magnon scattering processes. Further, the net magnetization arising from unequal, antiferromagnetically coupled sublattices in FIM architectures decouples magnetic excitations from dissipative electronic channels~\cite{kim2019low}, offering a robust route to minimize energy loss.

FIM insulating materials, including Li- and Al-containing ferrites and YIG (Y$_3$Fe$_5$O$_{12}$) possess low magnetic damping~\cite{omahoney2023aluminum, onbasli2014pulsed}. These materials exhibit interesting physical phenomena that contribute to the reduction of magnetic damping. Due to Fe\textsuperscript{3+} ions with electronic configuration (3$d^5$) along with half-filled 3\textit{d} orbitals, these Fe-based insulators exhibit strong magnetic exchange interactions and low spin-orbit coupling. The low spin-orbit coupling helps reduce the transfer of spin angular momentum to the lattice through phonons. The high-symmetry structure and, in some cases non-centrosymmetric structure help suppress magnetic anisotropy induced damping effects, allowing magnetization to precess with minimal energy loss. The relatively high exchange interactions, resulting in high magnetic ordering and usually low magnetic anisotropy, ensure stable magnetization precession. The minimized defects, which prevent the enhanced spin-orbit coupling usually from heavy elements, further contribute to the low damping. The large exchange stiffness and high coherence of spin waves, ensuring long magnon lifetimes prevent multi-magnon scattering. The lack of conduction electrons prevents losses from different induced effects such as spin Hall and spin pumping.

Due to the strongly correlated nature of the Fe-3\textit{d} states in Al-substituted Li-ferrites, a specialized treatment beyond standard density functional theory (DFT) is necessary ~\cite{adelstein2011structure, perdew1981self-interaction}. Conventional DFT functionals suffer significant self-interaction and delocalization errors, often failing to correctly predict the ground-state properties of these insulating materials~\cite{alvarez2023strongly}. To mitigate this, an onsite Coulomb interaction term (Hubbard \textit{U}) is generally added to the DFT Hamiltonian~\cite{dudarev1998electron, anisimov1991band}, however, calculating the value of $U$ and its validation inevitably reduces the predictive power of the method. Hybrid functionals \cite{heyd2003hybrid}, introducing a fraction of exact exchange, which go beyond such parameter dependent Hubbard model, improve the description of $3d$-electrons in these materials.

\begin{figure*}[t]
\centering
\includegraphics[width=0.92\textwidth]{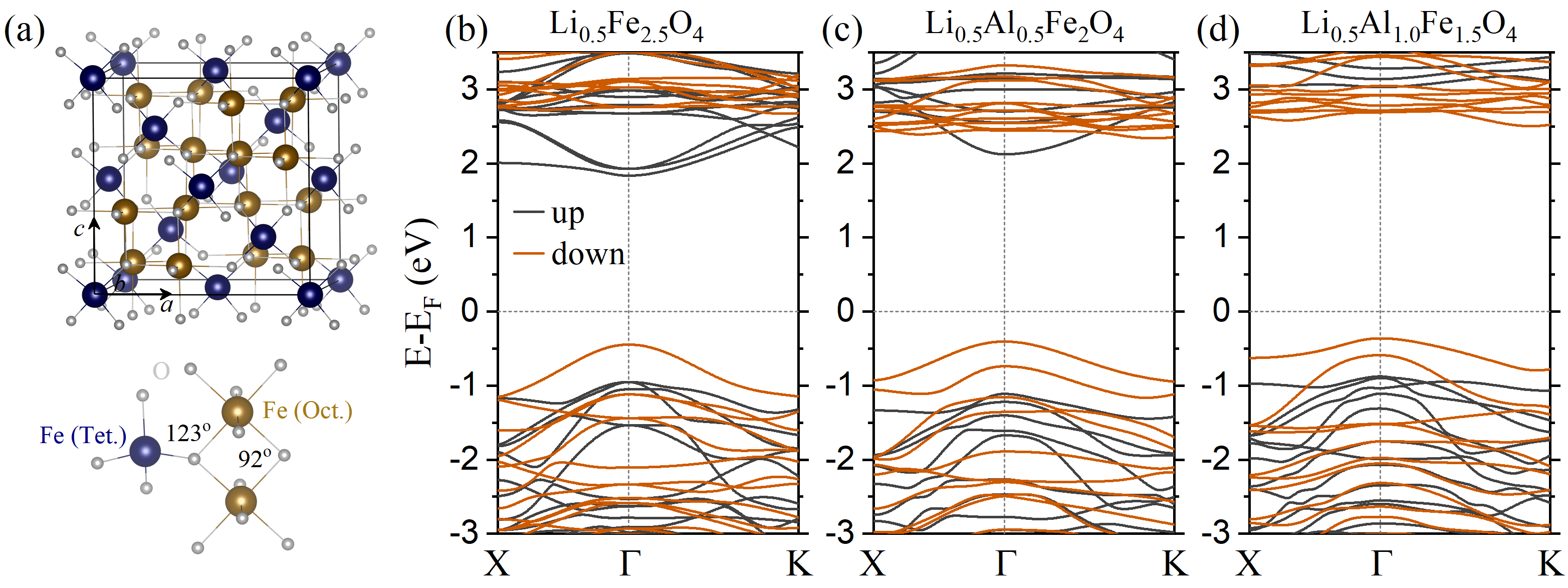}
\caption{(a) Crystal structure of Al-substituted Li-ferrite. There are two types of cation sites: tetrahedral and octahedral. The tetrahedral and octahedral Fe atoms form a 123$^o$ super-exchange angle in shared O atom, resulting in FIM super-exchange while the octahedral and octahedral Fe atoms form a 92$^o$ super-exchange angle in O atom, resulting in FM super-exchange. Spin projected band structures of (a)~$\text{Li}_{0.5}\text{Fe}_{2.5}\text{O}_4$, (b)~$\text{Li}_{0.5}\text{Al}_{0.5}\text{Fe}_{2.0}\text{O}_4$, and (c)~$\text{Li}_{0.5}\text{Al}_{1.0}\text{Fe}_{1.5}\text{O}_4$. The band gap evolves with the increase of Al concentration.}
\label{fig1}
\end{figure*}

Using advanced DFT calculations, this study reveals that Al and Li atoms preferentially occupy octahedral sites in spinel ferrite systems, a crucial material family for low-loss magnonics. Substituting Al into Li-ferrite transitions the material from a half-metallic ferrimagnetic ground state with in-plane anisotropy to an out-of-plane insulating state by removing the Fermi-level density of states. This structural and electronic modification enables ultra-low magnetic damping and significantly lowers the threshold for current-induced magnetization dynamics. Exchange interaction calculations confirm that strong FIM super-exchange occurs between oppositely aligned octahedral and tetrahedral Fe moments via mediating O atoms, producing highly stable, low-damping collective excitations suitable for scalable magnonic interconnects. Crucially, computed magnon spectra in this FIM system show a collective acoustic response reflecting overall iron site ratios alongside distinct, sub-lattice-selective optical modes localized on octahedral or tetrahedral networks, providing a microscopic link for site-specific chemical substitution effects on spin dynamics. Further, the structural straining effect strongly alters both low-frequency acoustic and intermediate-frequency optical phonon modes, driving Raman active modes with tailored spin-phonon coupling. The resulting unique magnon dispersions deviate significantly from standard ferromagnets and overlap closely with these distinct phonon branches; this intentional energetic alignment enables robust magnon-phonon coupling and exceptionally high cooperativity, providing the rapid energy exchange between magnetic spins and lattice vibrations necessary to realize advanced hybrid quantum magnonic platforms and high-efficiency magneto-elastic devices.

First-principles electronic structure calculations based on DFT were performed using the Vienna \textit{Ab initio} Simulation Package~\cite{furthmuller1996dimer, kresse1996efficiency} with projector augmented wave
pseudopotentials~\cite{blochl1994projector} and Perdew-Burke-Ernzerhof (PBE) functionals~\cite{perdew1996o} within the generalized gradient approximation. A plane wave energy cutoff of 500~eV, a smearing value of 0.1~eV, and a $\Gamma$-centered 8 $\times$  8 $\times$ 8 Monkhorst-Pack \textbf{k}-mesh were used in structural relaxation and electronic structure calculations. The self-consistent energy convergence criterion of 10$^{-6}$~eV is used in the calculations of the atomic and electronic structures, while the atomic positions and lattice constants are relaxed until the maximum force on each atom is less than 10$^{-4}$~eV/\AA. Spin-orbit coupling effects were incorporated into self-consistent field calculations. The underestimated band gaps obtained with PBE are corrected using hybrid functional calculations, which partially incorporate the exchange-correlation potential of standard DFT with the Hartree-Fock exchange~\cite{heyd2003hybrid}. Maximally localized Wannier functions, constructed from Fe $3d$ and O $2p$ orbitals, were used to describe the electronic structure and generate tight-binding Hamiltonians via the Wannier90 package~\cite{Pizzi2019s}. Subsequently, magnetic exchange interactions were evaluated using the \textsc{TB2J} code~\cite{he2021tb2j}. To investigate magnon dispersions, first we perform DFT+U with effective Hubbard parameter of 4 eV, validated by standard hybrid functional calculations and generate Wannier functions to construct a tight binding model from the wave functions. The resulting exchange parameters are then used to calculate magnon dispersions by diagonalizing the spin Hamiltonian within the linear spin-wave approximation via the Holstein--Primakoff transformation~\cite{holstein1940field}. 

The low damping spinel material forms in a cubic ($Fd\Bar{3}m$) structure with oxygen (O$^{2-}$ anions) making a closed-pack face-centered cubic lattice and Fe cations mostly with Fe$^{3+}$ occupying interstitial sites~\cite{fonin2005Surface}. The structure belongs to the high-symmetry $m\bar{3}m$ crystallographic point group and the $Fd\bar{3}m$ space group. Here, Fe atoms are distributed with octahedral and tetrahedral coordination (Fig. \ref{fig1}a). The structure may also be seen as layered sheets of tetrahedral and octahedral Fe atoms between closed-packed sheets of oxygen atoms. There are twice as many octahedral sites occupied as tetrahedral sites. To predict and understand the electronic and magnetic states of \ce{Li_{0.5}Al_{x}Fe_{2.5-x}O4}, we first analyzed the site preference energies of Li and Al at the octahedral and tetrahedral sites of Li-ferrite. The calculated electronic structure results show that both Al and Li prefer to occupy octahedral sites. Specifically, the octahedral site preference energy of Li (91.3~meV/atom) is higher than that of Al (58.0~meV/atom). While Al-ferrite (such as \ce{Al_{0.5}Fe_{2.5}O4}) exhibits half-metallic behavior and Li-ferrite (\ce{Li_{0.5}Fe_{2.5}O4}) behaves as an insulator, the Al-doped Li-ferrite retains an insulating FIM state. The calculated band gap of \ce{Li_{0.5}Al_{1.0}Fe_{1.5}O4} is 2.8~eV, which is 0.6~eV larger than the band gap of pure \ce{Li_{0.5}Fe_{2.5}O4}. Consequently, the band gap of \ce{Li_{0.5}Al_{0.5}Fe_{2.0}O4} falls between those of the $x=0$ and $x=1.0$ compositions. The direct band gap at $\Gamma$ from the valence band maximum (spin-down) to the conduction band minimum (spin-up) observed in \ce{Li_{0.5}Fe_{2.5}O4} is maintained in \ce{Li_{0.5}Al_{0.5}Fe_{2.0}O4} (Figs. \ref{fig1}b, \ref{fig1}c). However, this transitions into an indirect spin-down to spin-down band gap in \ce{Li_{0.5}Al_{1.0}Fe_{1.5}O4} (Fig. \ref{fig1}d). Aluminum substitution enhances the binding energy by pushing the occupied spin-up states toward lower energies and the unoccupied spin-up states toward higher energies. Interestingly, the band gap of \ce{Li_{0.5}Al_{1.0}Fe_{1.5}O4} compares well with that of YIG~\cite{Thiery2018Eelectrical}. The wide band gap of YIG is known to be advantageous for low-loss magnonics. Therefore, the wide band gap of \ce{Li_{0.5}Al_{1.0}Fe_{1.5}O4}---combined with its relatively simpler crystal structure, ease of synthesis, and straightforward device integration compared to YIG---makes it a highly promising candidate for low-loss magnonics.

Electronic structure calculations reveal a robust FIM ground state in Al-substituted Li-ferrites such as $\text{Li}_{0.5}\text{Fe}_{2.5}\text{O}_4$, $\text{Li}_{0.5}\text{Al}_{0.5}\text{Fe}_{2}\text{O}_4$ and $\text{Li}_{0.5}\text{Al}_{1.0}\text{Fe}_{1.5}\text{O}_4$ spinels, driven by oxygen-mediated super-exchange interactions between octahedral and tetrahedral Fe sites. In $\text{Li}_{0.5}\text{Fe}_{2.5}\text{O}_4$, lithium preferentially occupies octahedral sites alongside three $\text{Fe}^{3+}$ atoms, leaving two $\text{Fe}^{3+}$ atoms at tetrahedral sites with opposing magnetic moments of $4.14\,\mu_B/\text{Fe}$ ($O$) and $-4.03\,\mu_B/\text{Fe}$ ($T$). The $\text{Li}_{0.5}\text{Al}_{0.5}\text{Fe}_{2.0}\text{O}_4$ features two octahedral Fe, one octahedral Al, one octahedral Li, and two tetrahedral Fe atoms. Similarly, $\text{Li}_{0.5}\text{Al}_{1.0}\text{Fe}_{1.5}\text{O}_4$ features one octahedral Fe, two octahedral Al, one octahedral Li, and two tetrahedral Fe atoms. A dominant, antiparallel $\text{Fe}^O\text{-O-Fe}^T$ super-exchange yields large negative coupling energies of $-4.26\text{ meV}$ for $\text{Li}_{0.5}\text{Fe}_{2.5}\text{O}_4$, of $-4.9\text{ meV}$ for $\text{Li}_{0.5}\text{Al}_{0.5}\text{Fe}_{2}\text{O}_4$, and $-4.37\text{ meV}$ for $\text{Li}_{0.5}\text{Al}_{1.0}\text{Fe}_{1.5}\text{O}_4$, which significantly outweigh weaker intra-sub-lattice FM terms. In all compositions, the FIM coupling stems from directional orbital hybridization and bond angles that diverge from ideal $180^\circ$ Goodenough-Kanamori-Anderson (GKA) antiferromagnetism~\cite{rijn2022strain}. Spin-up Fe $3d$ states at tetrahedral sites hybridize with spin-up O $2p$ states, whereas spin-down Fe $3d$ states at octahedral sites couple with spin-down O $2p$ states through a $123^\circ$ bridging angle \cite{Tong2026Direct}.  Conversely, intra-sub-lattice $\text{Fe}^O\text{-O-Fe}^O$ interactions form a distorted $92^\circ$ angle near the $90^\circ$ threshold, inducing weak ferromagnetism. The primary $\text{Fe}^O\text{-O-Fe}^T$ super-exchange magnitude in Al-substituted Li-ferrites compares favorably with YIG~\cite{gorbatov2021magnetic}, establishing a parallel underlying magnetic mechanism.

\begin{figure}[!ht]
\centering	
\includegraphics[width=0.45\textwidth]{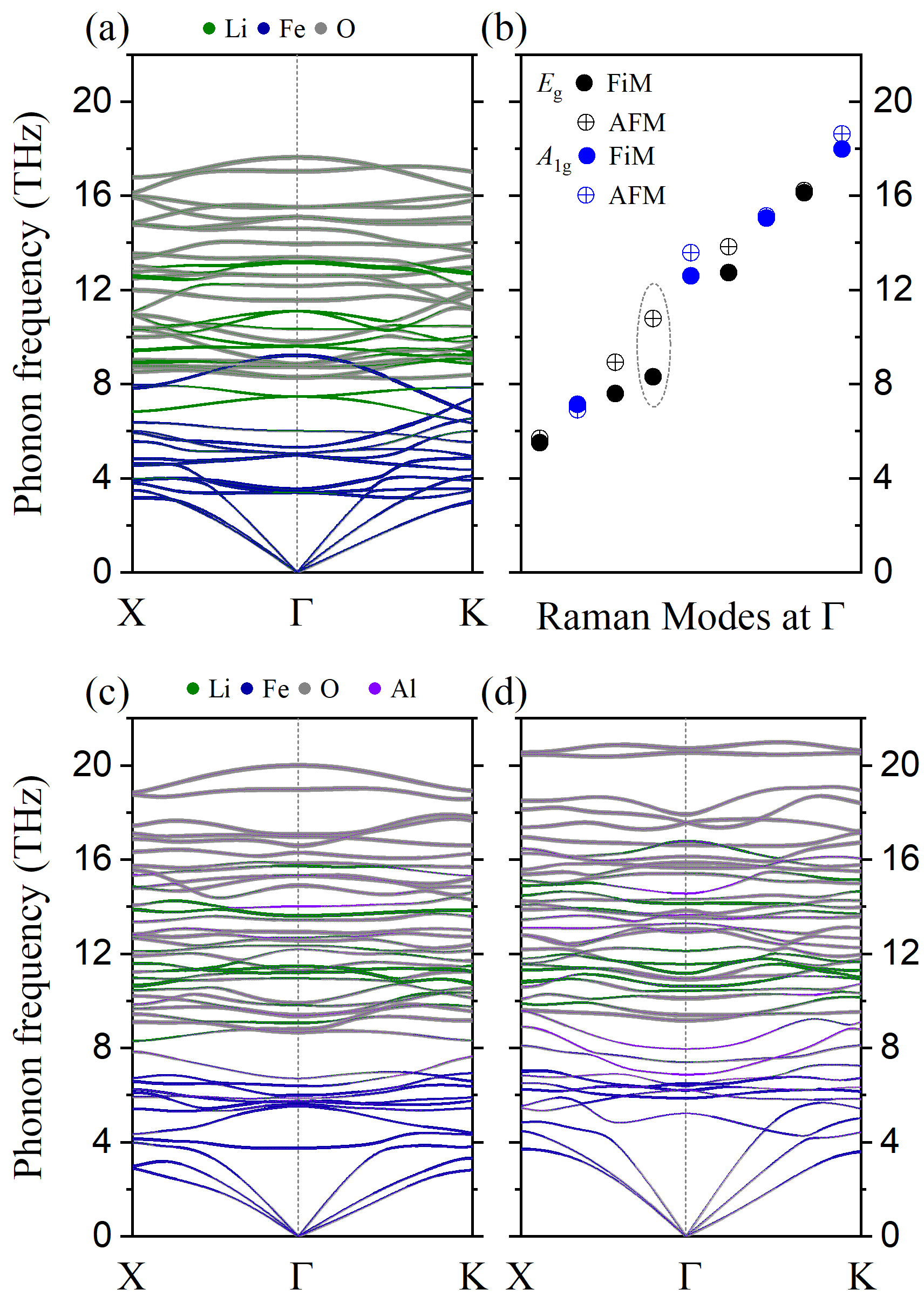}\hfill	
\caption{Atom resolved phonon dispersions and Raman-active frequencies from \textit{ab initio} calculations. FIM phonon dispersions for (a)~$\text{Li}_{0.5}\text{Fe}_{2.5}\text{O}_4$, (c)~$\text{Li}_{0.5}\text{Al}_{0.5}\text{Fe}_{2.0}\text{O}_4$, and (d)~$\text{Li}_{0.5}\text{Al}_{1.0}\text{Fe}_{1.5}\text{O}_4$; vertical dashed lines indicate high-symmetry points. (b)~Comparison of Raman-active frequencies between FIM and AFM configurations. The highlighted mode exhibits a spin-dependent shift of $\approx 2.5$~THz in~$\text{Li}_{0.5}\text{Fe}_{2.5}\text{O}_4$}
\label{fig3:phonon_raman}
\end{figure}

To elucidate the magnetic behavior of Al-substituted Li-ferrites, we investigate their magnetocrystalline anisotropy energy (MAE) via total energy calculations. We define this energy as $\Delta E_{\text{MAE}} = E_{110} - E_{001}$, representing the difference in total energy when the magnetization is aligned along the $[110]$ versus the $[001]$ crystallographic directions. Under this definition, a positive value establishes the $[001]$ axis as the energetically favored easy axis. For pristine $\text{Li}_{0.5}\text{Fe}_{2.5}\text{O}_4$, calculations yield a robust positive MAE of $85~\mu\text{eV}$, establishing a definitive perpendicular magnetic ground state along the $[001]$ easy axis. Upon systematic Al substitution, the MAE exhibits a distinct non-monotonic evolution, dropping to $42~\mu\text{eV}$ for $\text{Li}_{0.5}\text{Al}_{0.5}\text{Fe}_{2.0}\text{O}_4$ before slightly recovering to $47~\mu\text{eV}$ in $\text{Li}_{0.5}\text{Al}_{1.0}\text{Fe}_{1.5}\text{O}_4$. While high-spin $\text{Fe}^{3+}$ ($d^5$) ions possess quenched orbital angular momentum, local geometric distortions break the cubic symmetry within the tetrahedral and octahedral sublattices. This asymmetry hybridizes the $\text{Fe}~3d$ states with $\text{O}~2p$ ligands, driving the valence electronic charge density away from sphericity and pinning the easy axis via spin-orbit coupling. The non-monotonic trend under non-magnetic $\text{Al}^{3+}$ dilution reflects a complex structural interplay and partial cancellation between the competing tetrahedral and octahedral site anisotropy. Unlike traditional low-loss magnonic materials like YIG that require external strain tuning for out-of-plane alignment \cite{Ke2024Intrinsic}, Al-substituted Li-ferrites possess intrinsic perpendicular magnetic anisotropy. This native alignment enhances spin-wave stability and restructures magnon dispersions. 

To elucidate the intrinsic relationship between lattice dynamics and Raman-active vibrational modes, we systematically calculate and analyze the atom resolved phonon dispersions of cubic inverse spinel $\mathrm{Li}_{0.5}\mathrm{Fe}_{2.5}\mathrm{O}_{4}$ (Figs.~\ref{fig3:phonon_raman}a, \ref{fig3:phonon_raman}c, \ref{fig3:phonon_raman}d). The complete absence of imaginary frequencies across the entire Brillouin zone confirms the rigorous structural and dynamical stability of this phase. While the acoustic branches exhibit conventional linear dispersion relations originating from the zone center ($\Gamma$) and dictate the low-temperature thermal transport, the optical branches display notably flat dispersion profiles throughout the upper vibrational spectrum. These flat profiles signify highly localized atomic vibrations, corresponding to a suppression of phonon group velocities and indicating restricted energy propagation between adjacent unit cells. 

Within this structural framework, the cations occupy two highly distinct local environments: the tetrahedral sites exhibiting $\bar{4}3m$ local site symmetry, and the octahedral sites possessing $\bar{3}m$ local site symmetry. The local lowering of the lattice symmetry breaks down foundational cubic degeneracies. Group theoretical analysis indicates that this reduction lifts structural degeneracies to yield a complex vibrational spectrum containing $4A_{1g}$ and $5E_g$ non-triplet Raman-active symmetry types. The four $A_{1g}$ modes represent highly symmetric, isotropic breathing and stretching displacements of the oxygen coordination cages driven by high-frequency tetrahedral ($\text{FeO}_4$) sublattices. Conversely, the five $E_g$ modes denote doubly-degenerate symmetric bending and twisting vibrations acting primarily within the octahedral blocks of ordered Li and Fe ions without shifting the structural center of mass. Beyond these active optical branches, acoustic phonon modes and silent symmetries round out the comprehensive dispersion curves, directly establishing the baseline thermal and magneto-elastic characteristics of the material.

The calculated phonon spectrum is distinct into three energetic regimes: a low-frequency region (0 to $\sim$4~THz), an intermediate-frequency band ($\sim$4 to $\sim$13~THz), and a high-frequency domain extending above $\sim$13~THz. Atom-resolved dispersions reveal that the low-frequency modes stem from a coupled mixture of vibrations dominated by the heavier Fe cations executing rigid polyhedral translations. Conversely, the intermediate and high-frequency regions are governed by the interlaced dynamics of Li, Fe, and light O atoms. Crucially, oxygen-driven vibrations span the entire spectrum, serving as the core framework for both the tetrahedral $\text{FeO}_{4}$ and octahedral $\text{FeO}_6$ local environments. This structural arrangement induces strong orbital hybridization and collective oscillations among adjacent polyhedra, yielding highly mixed vibrational modes---a behavior reminiscent of complex polyhedral networks observed in cerium vanadates and phosphates~\cite{Paudyal2025Singly}. This intricate coupling between local polyhedral distortions and extended lattice dynamics establishes the fundamental microscopic origins that dictate the selection rules, frequencies, and intensities of the corresponding zone-center Raman-active modes.

To quantify how long-range magnetic order influences these lattice dynamics, we evaluate the spin--phonon coupling strength associated with the system's Raman-active modes by comparing the zone-center frequencies across distinct magnetic configurations. In the magnetically ordered phase, the shift in phonon frequency serves as a direct quantitative measure of spin-lattice interactions, revealing a profound correlation between atomic vibrations and localized electron spins. This interaction is described phenomenologically using a static spin-spin correlation function, where the spin-dependent frequency shift follows $\Delta\omega = \lambda \langle \mathbf{S}_i \cdot \mathbf{S}_j \rangle$, where $\lambda$ denotes the spin-phonon coupling constant. Optical Raman-active modes at the $\Gamma$ point of the Brillouin zone exhibit pronounced modifications---most notably in the low-frequency regime---depending directly on the underlying alignment of the magnetic sublattices (Fig.~\ref{fig3:phonon_raman}b). By incorporating an iron ($\text{Fe}^{3+}$) high-spin state alongside the spin-spin correlation function, the maximum observed change in phonon frequency ($\Delta\omega \approx 2.5$~THz) yields an extracted coupling constant $\lambda \approx 0.4$~THz. This value is comparatively smaller than those reported in many other complex transition metal oxides~\cite{calder2015enhanced, zhang2021first, kim2020spin, Paudyal2025Singly}. This finding confirms Li-ferrite as a hallmark material for low magnetic damping, indicating that while the lattice dynamics are cleanly modulated by magnetic ordering, the interactions remain moderate enough to avoid severe intrinsic scattering.

Aluminum substitution systematically blue-shifts these phonon frequencies, as directly evidenced by progressing from the $\text{Li}_{0.5}\text{Al}_{0.5}\text{Fe}_{2.0}\text{O}_4$ ($x=0.5$) to the more heavily substituted $\text{Li}_{0.5}\text{Al}_{1.0}\text{Fe}_{1.5}\text{O}_4$ ($x=1.0$) stoichiometry. When lighter $\text{Al}^{3+}$ ions increasingly replace heavier $\text{Fe}^{3+}$ ions---primarily at the octahedral sites of the inverse spinel structure---the overall reduced mass of the vibrating clusters decreases, pushing the characteristic vibrational modes to higher frequencies. Furthermore, because the ionic radius of $\text{Al}^{3+}$ is smaller than that of $\text{Fe}^{3+}$, this compositional shift systematically contracts the global lattice parameters and shortens the local metal-oxygen bond lengths. This structural compression stiffens the interatomic bonds and increases the respective force constants, causing a distinct, concentration-dependent migration of the infrared-active and Raman-active bands toward higher wavenumbers. Ultimately, modulating the lattice dynamics between these specific compositions provides a highly controllable framework for tuning magnon-phonon coupling. This degree of structural freedom is critical for minimizing spin-lattice relaxation bottlenecks, preventing premature magnon decay, and optimizing high-frequency spintronic devices.

\begin{figure}[!ht]
    \centering
    \includegraphics[width=0.49\textwidth]{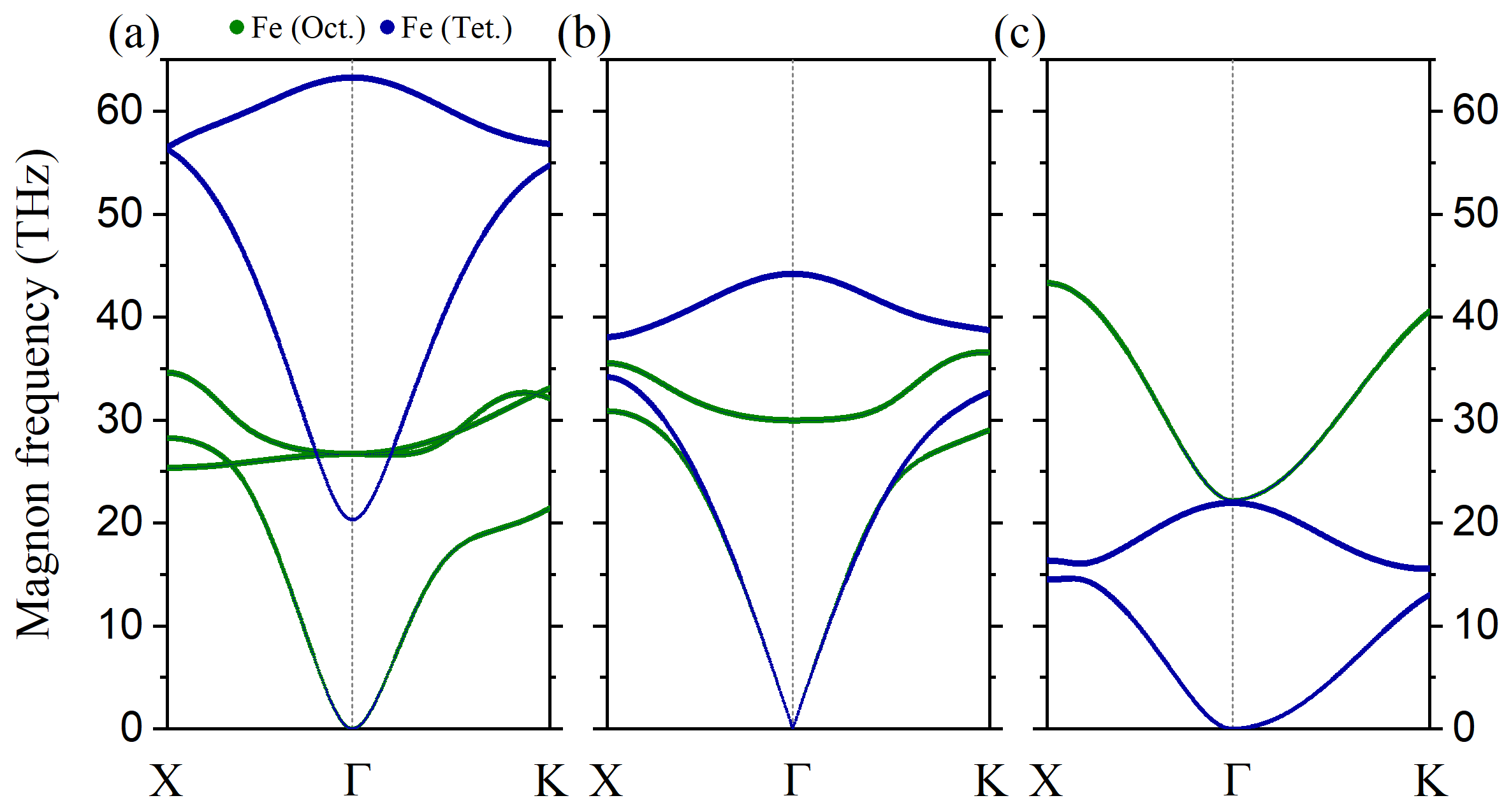}
    \includegraphics[width=0.35\textwidth]{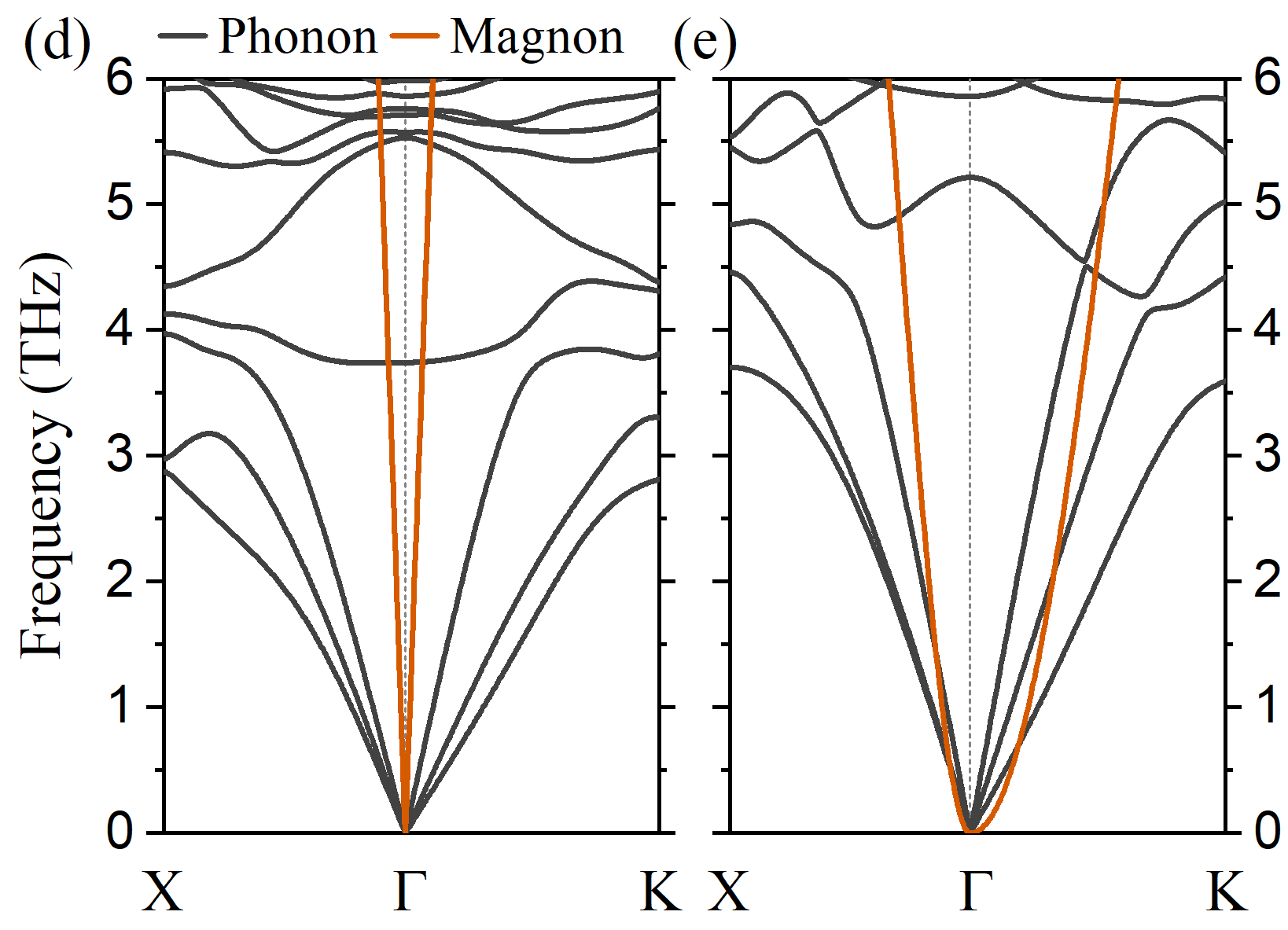}
    \caption{Sub-lattice-resolved magnon dispersions for (a) $\text{Li}_{0.5}\text{Fe}_{2.5}\text{O}_4$, (b) $\text{Li}_{0.5}\text{Al}_{0.5}\text{Fe}_{2.0}\text{O}_4$, and (c) $\text{Li}_{0.5}\text{Al}_{1.0}\text{Fe}_{1.5}\text{O}_4$. Increasing Al substitution reduces the number of magnetic Fe sites per primitive cell from five to four and three, respectively. This decrease shrinks the number of magnon branches and significantly reconstructs the spectrum. The low energy magnon and phonon modes are shown together in $\text{Li}_{0.5}\text{Al}_{0.5}\text{Fe}_{2.0}\text{O}_4$ (d) and $\text{Li}_{0.5}\text{Al}_{1.0}\text{Fe}_{1.5}\text{O}_4$(d) (e).}
    \label{fig4:magnon_dispersion}
\end{figure}

Chemical tuning provides a powerful route to engineering magnetic excitations in quantum 
materials, yet isolating these spin dynamics from concurrent structural modifications remains 
an elusive challenge. Here, we demonstrate a microscopic mechanism for magnon engineering 
in inverse-spinel ferrimagnets using lithium ferrite ($\mathrm{Li}_{0.5}\mathrm{Fe}_{2.5}\mathrm{O}_{4}$) 
and its aluminum-substituted derivatives ($\mathrm{Li}_{0.5}\mathrm{Al}_{0.5}\mathrm{Fe}_{2.0}\mathrm{O}_{4}$ 
and $\mathrm{Li}_{0.5}\mathrm{Al}_{1.0}\mathrm{Fe}_{1.5}\mathrm{O}_{4}$) as a model system. 
Linear spin-wave theory calculations, utilizing interatomic exchange parameters extracted via DFT, map the evolution of the collective magnetic 
spectra along a unified $X$--$\Gamma$--$K$ high-symmetry path (Fig.~\ref{fig4:magnon_dispersion}). 
By referencing this path strictly to the Brillouin zone of the parent cubic structure, 
we effectively isolate electronic and magnetic perturbations from confounding structural 
cell transformations. In the parent compound, the antiparallel alignment of unequal 
tetrahedral and octahedral sub-lattice moments yields a characteristic FIM ground 
state that manifests as high-energy optical branches alongside a gapless acoustic mode. 
This ferromagnet-like Goldstone mode exhibits a quadratic long-wavelength dispersion at the $\Gamma$ point, resulting directly from 
the spontaneous breaking of continuous spin-rotation symmetry under an isotropic 
Heisenberg exchange Hamiltonian. In contrast, \(\mathrm{Li}_{0.5}\mathrm{Al}_{0.5}\mathrm{Fe}_{2.0}\mathrm{O}_{4}\) exhibits nearly linear acoustic modes emerging from the zone center (Fig. \ref{fig4:magnon_dispersion}b). This behavior stems from its low net magnetic moment, which arises from an equal distribution of oppositely aligned octahedral and tetrahedral Fe sites.  Crucially, a close inspection of the corresponding 
phonon and magnon spectra reveals a low-frequency magnon-phonon intersection in both the Al-substituted compounds (Figs. \ref{fig4:magnon_dispersion}d, \ref{fig4:magnon_dispersion}e). Higher degree of magnon-phonon intersection is found in $\mathrm{Li}_{0.5}\mathrm{Al}_{1.0}\mathrm{Fe}_{1.5}\mathrm{O}_{4}$ as compared to that in $\mathrm{Li}_{0.5}\mathrm{Al}_{0.5}\mathrm{Fe}_{2.0}\mathrm{O}_{4}$. Driven by magneto-elastic coupling, this interaction mixes their respective degrees of freedom to form hybrid quasi-particles; rather than consistently opening a distinct anti-crossing energy gap, this hybridization primarily manifests as a dramatic enhancement of scattering intensity at the intersection points, thereby significantly boosting spin transport.

Our calculations reveal that systematic substitution of nonmagnetic $\mathrm{Al}^{3+}$ ions 
into the octahedral sites drives a dramatic, non-uniform restructuring of these collective spin dynamics rather than a simple, rigid down-scaling of the parent spectrum. As the aluminum concentration ($x$) increases, the number of active magnon branches drops from five at $x=0$ (corresponding to the five magnetic $\mathrm{Fe}$ sites in the primitive cell) to 
four at $x=0.5$, and finally to three at $x=1.0$. Concurrently, the maximum optical 
excitation bandwidth undergoes a profound suppression from $63$~THz to $44$~THz 
and $42.4$~THz, respectively. This softening originates from a highly selective magnetic 
reduction of the exchange interactions including inter-sub-lattice and intra-sub-lattice. While the dominant inter-sub-lattice exchange 
$J_{\mathrm{Fe}^{O}\!-\mathrm{Fe}^{T}}$ is protected by a robust 
$\mathrm{Fe}^{O}\!-\mathrm{O}-\mathrm{Fe}^{T}$ local bond geometry near $123^{\circ}$, the 
depletion of magnetic octahedral iron pairs drastically curtails the intra-sub-lattice 
FM exchange $J_{\mathrm{Fe}^{O}\!-\mathrm{Fe}^{O}}$. This targeted suppression 
lowers the effective exchange stiffness, simultaneously flattening the acoustic curvature and shifting the optical precessions to lower energies (Fig. \ref{fig4:magnon_dispersion}c).

To map the exact spatial distribution of these collective excitations, we compute the 
sub-lattice-resolved magnon spectra, which reveal a pronounced separation of the spin-wave 
modes dictated by the local Fe coordination. At the zone center ($\Gamma$), the acoustic 
Goldstone mode exhibits weight fractions of $W_{\rm oct}=0.60$ and $W_{\rm tet}=0.40$, 
characterized by a uniform, phase-coherent participation of all five Fe sites within the 
primitive cell of $\mathrm{Li}_{0.5}\mathrm{Fe}_{2.5}\mathrm{O}_{4}$. This exact $3:2$ ratio is not merely accidental; it directly reflects the 
structural stoichiometry of the octahedral and tetrahedral Fe populations, confirming 
a rigid, collective rotation of the ordered FIM ground state. 
In stark contrast, the high-energy optical modes break away from this collective behavior 
to exhibit extreme sub-lattice selectivity. The nearly degenerate optical branches residing 
at $\sim26.7$~THz are found to be almost entirely localized within the octahedral Fe network, 
meaning their dynamical amplitudes drop to near-zero on the tetrahedral sites. Conversely, 
the apex optical mode at $\sim63.2$~THz displays a purely tetrahedral character, leaving 
the octahedral network entirely stationary. An intermediate optical branch at $\sim20.3$~THz 
remains strongly hybridized, maintaining an approximate $40\%$ octahedral and $60\%$ 
tetrahedral character. Substitution with \(\mathrm{Al}^{3+}\) shifts the tetrahedral \(\mathrm{Fe}\) magnon modes to lower energies, ultimately transforming them into acoustic modes in \(\mathrm{Li}_{0.5}\mathrm{Al}_{1.0}\mathrm{Fe}_{1.5}\mathrm{O}_{4}\).

This stark dynamical segregation demonstrates that despite the strong inter-sub-lattice exchange 
coupling tying the system together, the high-energy optical spectrum preserves a remarkably 
distinct sub-lattice identity. Such a separation provides an explicit microscopic framework 
for understanding how targeted chemical substitutions can selectively reconstruct specific 
regions of the magnon spectrum without destabilizing the global magnetic order. In particular, 
nonmagnetic $\mathrm{Al}^{3+}$ substitution does not act as a spatial average; it modifies 
both the magnetic-site population and the local exchange pathways unevenly. As a consequence, 
octhedral-, tetrahedral-, and mixed-sublattice magnons do not evolve uniformly with doping, 
enabling a highly localized, mode-specific manipulation of the spin-wave spectrum. While the 
collective acoustic branch remains protected at long wavelengths---governing the macroscopically 
probed low-loss magnonic transport---the segregated optical modes behave as distinct microscopic 
channels. These high-frequency channels allow local chemical disorder, defects, and specific 
lattice phonon modes to couple selectively into individual spin sublattices, unlocking an 
atomic-scale lever to tune magnon lifetimes.

Ultimately, the concurrent evolution of the spin-wave stiffness and these highly selective optical modes yields a cohesive framework for predicting both the thermodynamic and symmetry-breaking properties of complex ferrimagnets. At low temperatures, doping-induced modifications to the spin-wave stiffness fundamentally alter thermally excited magnons, establishing a direct link between atomic-scale connectivity and the macroscopic Bloch correction to the magnetization. At elevated temperatures, the downshifted optical branches become thermally populated, introducing distinct thermodynamic corrections that stem entirely from out-of-phase sub-lattice precessions. Furthermore, realistic experimental conditions—such as spin-orbit coupling, local structural distortions, or chemical disorder—inevitably break continuous rotational symmetry. This symmetry breaking opens an anisotropy-induced zone-center gap, modifying the acoustic dispersion to \(\hbar\omega_{\mathrm{ac}}(k)\simeq\Delta+Dk^{2}\). Tracking the co-evolution of spin-wave stiffness ($D(x)$) and anisotropy-induced zone-center gap ($\Delta(x)$) and the optical gaps through eigenvector calculations reveals that \(\mathrm{Al}^{3+}\) incorporation acts as a precise microscopic control parameter. This chemical intervention systematically dismantles the connectivity of the octahedral network while preserving the core inter-sub-lattice pathways, thereby bridging atomic-scale cation ordering with experimentally accessible spin dynamics.

In summary, through first-principles electronic structure and exchange interaction calculations, we establish Al-substituted Li-ferrite as an insulating FIM material characterized by highly stable super-exchange interactions and a completely eliminated DOS at the Fermi level. Calculations reveal that preferential octahedral site occupation by Al and Li induces an inherent, composition-driven lattice strain. This microscopic distortion directly mandates a transition from an in-plane to an out-of-plane magnetocrystalline anisotropy, providing a robust mechanism to drastically lower the threshold for current-induced magnetization switching. Simultaneously, this intrinsic straining effect profoundly alters the low-frequency acoustic and optical phonon branches, suppressing spin-phonon scattering to yield exceptionally low-loss magnonic excitations. Crucially, the magnon spectra in this FIM system show a collective acoustic response reflecting overall iron site ratios alongside distinct, sub-lattice-selective optical modes localized on octahedral or tetrahedral networks, providing a microscopic link for site-specific chemical substitution effects on spin dynamics. These unconventional magnon dispersions strongly intersect with lattice vibrations, driving a robust, high-cooperativity magnon-phonon coupling. By bridging fundamental quantum mechanical interactions with macroscopic magneto-elastic behaviors, these insights demonstrate that Al-substituted spinel Li-ferrites represent an exceptional, multi-functional material class for low-loss magnonics. Ultimately, these findings unlock a predictive design paradigm for next-generation quantum materials, wherein coupled spin-lattice degrees of freedom are deterministically engineered for coherent magnon control.

This work is supported as part of the Center for Energy Efficient Magnonics, an Energy Frontier Research Center funded by the U.S. Department of Energy, Office of Science, Basic Energy Sciences, under Award number DE-AC02-76SF00515. H.P and D.P acknowledges the use of the computational facilities on the Frontera supercomputer at the Texas Advanced Computing Center (TACC) via the pathway allocation, DMR23051.

\bibliography{references}
\end{document}